# Emergence of ferromagnetism at the onset of moiré Kondo breakdown

Wenjin Zhao[1*], Bowen Shen[2*], Zui Tao[2*], Sunghoon Kim[3], Patrick Knüppel[3], Zhongdong Han[3], Yichi Zhang[3], Kenji Watanabe[4], Takashi Taniguchi[4], Debanjan Chowdhury[3], Jie Shan[1,2,3**], Kin Fai Mak[1,2,3**]

[1]Kavli Institute at Cornell for Nanoscale Science, Ithaca, NY, USA
[2]School of Applied and Engineering Physics, Cornell University, Ithaca, NY, USA
[3]Laboratory of Atomic and Solid State Physics, Cornell University, Ithaca, NY, USA
[4]National Institute for Materials Science, 1-1 Namiki, 305-0044 Tsukuba, Japan

*These authors contributed equally.
**Email: jie.shan@cornell.edu; kinfai.mak@cornell.edu

**The interaction of a lattice of localized magnetic moments with a sea of conduction electrons in Kondo lattice models induces rich quantum phases of matter, such as Fermi liquids with heavily renormalized electronic quasiparticles, quantum critical non-Fermi liquid metals and unconventional superconductors, among others [1–4]. The recent demonstration of moiré Kondo lattices has opened the door to investigate the Kondo problem with continuously tunable parameters [5–10]. Although a heavy Fermi liquid phase has been identified in moiré Kondo lattices, the magnetic phases and Kondo breakdown transitions remain unexplored. Here we report a density-tuned Kondo destruction in AB-stacked $MoTe_2/WSe_2$ moiré bilayers by combining magneto transport and optical studies. As the itinerant carrier density decreases, the Kondo temperature decreases. At a critical density, we observe a heavy Fermi liquid to insulator transition, and a nearly concomitant emergence of ferromagnetic order. The observation is consistent with the scenario of a ferromagnetic Anderson insulator and suppression of the Kondo screening effect. Our results pave the path for inducing other exotic quantum phase transitions in moiré Kondo lattices [6–15].**

**Main**

Moiré materials provide a highly tunable physical platform for strongly correlated electron phenomena in two dimensions [16–19]. When a flat moiré electronic band is half-filled, the strong on-site Coulomb repulsion can localize electrons on the moiré lattice to form a Mott insulator, leading to a filled lower Hubbard band (LHB) and an empty upper Hubbard band (UHB) that are separated by a Mott gap $\sim U$ (Ref. [20–22]). In the absence of charge fluctuations, this leads to a lattice of well-formed local magnetic moments. When they exchange-couple to itinerant electrons from a dispersive band (conveniently from a different layer of the moiré structure), a Kondo lattice can be constructed [1–10] (Fig. 1a-c). Since the local moments and the itinerant electrons reside in different layers, both the interactions and the itinerant electron density can be continuously gate-tuned to explore the many-body quantum phase diagram [5–10].

A moiré Kondo lattice has been experimentally demonstrated in hole-doped AB-stacked $MoTe_2/WSe_2$ moiré bilayers [5]. The lattice mismatch between the two transition metal dichalcogenide layers produces a triangular moiré lattice with period $a_M \approx 5$ nm and density $n_M \approx 5 \times 10^{12}$ cm$^{-2}$ (Ref. [5,23–25]). The Wannier states of the topmost valence band of the MoTe$_2$ (Mo) and WSe$_2$ (W) layers are centered at two distinct high-symmetry stacking sites and constitute the two sublattices of a staggered honeycomb lattice [26–29] (Fig. 1b). The two layers are weakly coupled because interlayer tunneling is spin-suppressed in the AB stacking structure. Furthermore, the two layers experience moiré potentials of different strength, resulting in a relatively flat Mo band and dispersive W band [5,23,26–29]. With two gates (Fig. 1a), one can introduce one hole per moiré unit cell in the Mo layer, which becomes a Mott insulator, and place the dispersive W band between the two Hubbard bands with $x$ holes per unit cell (Fig. 1c). This is effectively a charge transfer insulator with a charge transfer gap $\Delta$ between the Fermi energy and the LHB. The structure has been shown to realize a Kondo lattice model [5]. A heavy Fermi liquid phase has been identified, but the magnetic phases and the nature of the Kondo-breakdown transition remain unknown.

In this study, we demonstrate a density-tuned Kondo breakdown in the MoTe$_2$/WSe$_2$ moiré Kondo lattice. As the itinerant hole density $x$ decreases, we observe a heavy Fermi liquid to insulator transition at critical density $x_c \approx 0.04$. From the temperature dependence of the magnetic susceptibility at high temperature, we identify ferromagnetic correlations between the local moments for a broad range of doping $x \lesssim 0.1$. But magnetic ordering starts to emerge at a lower density near the onset of Kondo breakdown revealed by the anomalous Hall resistance and magnetic hysteresis. The observed Kondo breakdown transition and emergence of a ferromagnetic insulator are distinct from the standard Doniach phase diagram [14] and earlier experiments reporting ferromagnetic metals and Kondo breakdown as a function of pressure and other tuning parameters [30,31].

**Metal-insulator transition**
We first identify the reported Kondo lattice regime in AB-stacked MoTe$_2$/WSe$_2$ moiré bilayers by performing transport measurements [5]. Dual-gated Hall bar devices of the MoTe$_2$/WSe$_2$ moiré are fabricated by the layer-by-layer transfer method (Methods). Figure 1d shows the longitudinal resistance, $R_{xx}$, as a function of hole density ($\nu$ in units of $n_M$) in the bilayer and out-of-plane electric field ($E$) at 10 mK and zero magnetic field. The values of $\nu$ and $E$ are determined from the two gate voltages and the independently calibrated gate capacitances. They control the Fermi level and the charge transfer gap $\Delta$ in the Kondo lattice regime as shown in Fig. 1c. The grey area corresponds to the parameter space that cannot be accessed either due to the limited gate voltages or the large sample/contact resistance. The region with near-zero resistance around $\nu = 1$ and high electric fields ($E > 0.645$ V/nm) corresponds to the reported quantum anomalous Hall state [24], where the Mo and W states strongly hybridize. The region between the two red dashed lines (at lower electric fields and $\nu = 1 + x$) is the Kondo lattice regime, identified from the electric-field independent Landau levels formed by holes in the W layer; and the Mott gap $U$ is estimated to be 22 meV (Extended Data Fig. 1).

Figure 1e shows the temperature ($T$) dependence of $R_{xx}$ at varying itinerant hole densities. We focus on the line cut along $E \approx 0.638$ V/nm (black dashed line, Fig. 1d). Similar results are observed at other electric-field cuts (Extended Data Fig. 2 illustrating the result at $E = 0.634$ V/nm). At large $x$, the resistance displays a peak around temperature $T^*$, below which $R_{xx}$ scales quadratically with temperature, $R_{xx} = R_0 + AT^2$, with coefficient $A$. This is consistent with the reported heavy Fermi liquid behavior [5] with $T^*$ denoting the characteristic Kondo temperature, below which coherent metallic-like transport develops, and $R_0$ representing the residual resistance limited by elastic impurity scattering. Across a number of heavy fermion compounds, the coefficient $A$ tracks the evolution of the quasiparticle effective mass according to the Kadowaki-Woods scaling [32]. Here it is more than an order of magnitude larger than at $\nu = 2 + x$ where the local moments are absent (Extended Data Fig. 3a), also in agreement with the earlier report [5]. At small $x$, we observe an insulating behavior with diverging resistance at low temperatures. The critical point of the metal-insulator transition ($x_c \approx 0.04$) is determined by following the criterion that $R_{xx}$ exhibits a power-law temperature dependence at the critical point [33] (Extended Data Fig. 4a). The value of the critical density is weakly sample dependent. (The results from an additional device are shown in Extended Data Fig. 5.)

**Magnetic correlations**
We probe the correlation between local moments across the metal-to-insulator transition by performing the reflective magneto circular dichroism (MCD) spectroscopy [17,22] (Methods). Specifically, we integrate the MCD spectrum over the exciton resonance of MoTe$_2$ (Extended Data Fig. 6). As shown by earlier studies, the integrated MCD (or MCD for short) is proportional to the valley polarization or the local moment magnetization in the Mo layer [22,34]. Figure 2a shows MCD as a function of out-of-plane magnetic field for representative hole densities at $E = 0.638$ V/nm (same as above) and $T = 1.6$ K. At $x = 0.03$, the MCD displays a steep slope around zero field and rapidly saturates with increasing field. At higher densities, the zero-field slope decreases, and concurrently, saturation occurs at higher magnetic fields. A similar response is observed at lower densities.

The density dependence of the magnetic susceptibility, $\chi$, is illustrated in Fig. 2b, where $\chi$ is extracted from the zero-field slope of the MCD [22]. The magnetic susceptibility peaks around $x = 0.03$ (slightly below $x_c$) and falls off on both sides. Further, both the large susceptibility and the pronounced peak are quenched upon raising the sample temperature to a few kelvins.

We analyze the temperature dependence of the susceptibility in Fig. 2c. For all densities, the high-temperature ($T > \theta$) susceptibility follows the Curie-Weiss dependence, $\chi^{-1} \propto T - \theta$ (solid lines), where the Curie-Weiss temperature $\theta$ reflects the interaction energy scale of the local moments. The Curie-Weiss temperature is slightly negative at $x = 0$ (bottom panel), suggesting a weak antiferromagnetic exchange between local moments in the charge transfer insulator. It becomes positive with increasing density ($x = 0.03$, middle panel), suggesting the emergence of ferromagnetic correlations. Further at $x = 0.1$ (top panel), $\theta$ is again slightly negative and the correlation is antiferromagnetic/paramagnetic. In this case, a heavy Fermi liquid emerges below the Kondo temperature denoted by the arrow, and the magnetic susceptibility is nearly a constant, which is characteristic of the Pauli-like behavior in a Fermi liquid [4].

The doping dependence of the Curie-Weiss temperature is summarized in Fig. 4. The vertical dashed line marks the critical density for the metal-insulator transition. The magnetic correlations between the local moments are antiferromagnetic ($\theta < 0$) around $x = 0$ and $x \gtrsim 0.1$, and the magnitude of $\theta$ increases with $x$ for $x \gtrsim 0.1$. The correlations are ferromagnetic ($\theta > 0$) between $x = 0$ and $0.1$; $\theta$ is peaked at a density near $x_c$. The doping dependence of $\theta$ and $\chi$ at 1.6 K (Fig. 2b) is fully consistent. Specifically, local moments with ferromagnetic (antiferromagnetic) correlations are highly susceptible (resistant) to small magnetic fields at low temperatures. The relatively high temperature of the MCD measurements, however, does not allow us to observe magnetic ordering.

**Magnetic ordering**
Below we perform magneto transport measurements down to 10 mK to probe long-range magnetic ordering. Figure 3a shows the magnetic-field dependence of Hall resistance ($R_{xy}$) for varying hole densities at 10 mK (doping below $x = 0.03$ cannot be accessed by the magneto transport measurement due to the large sample resistance). At $x = 0.07$, we observe a very small Hall response, which scales linearly with magnetic field. For smaller $x$, in addition to the ordinary Hall response, we observe an anomalous Hall response manifested as a resistance jump ($R_{xy}^{(a)}$) near zero field. In contrast, the anomalous Hall response is absent at $\nu = 2 + x$ for any $x$, where the local moments are absent (Extended Data Fig. 3b).

Figure 3b shows $R_{xy}$ at $x = 0.03$ for both forward and reverse magnetic-field sweep directions. At 10 mK, we observe a sharp anomalous Hall response with a small saturation field (about 3 mT) and magnetic hysteresis. As temperature increases, $R_{xy}$ decreases and the anomalous Hall response broadens. Above 2 K, hysteresis is no longer discernible. The sharp anomalous Hall response, together with the magnetic hysteresis at low temperatures, supports the emergence of ferromagnetic order.

We examine the magnetic ordering temperature, $T_c$, by studying $R_{xy}^{(a)}$ as a function of doping density and temperature (Fig. 3c). As $x$ increases, $R_{xy}^{(a)}$ decreases and continuously vanishes. As $T$ increases to a few kelvins, $R_{xy}^{(a)}$ vanishes for all doping densities. We estimate the temperature scale for ordering using the value at which $R_{xy}^{(a)}$ drops to 10% of its value at 10 mK (Extended Data Fig. 7). The values of $T_c$ are included in Fig. 4.

**Discussions**
The phase diagram in Fig. 4 summarizes the main findings of this study. The state at $x = 0$ represents a charge-transfer insulator with weak antiferromagnetic correlations between the local moments. Introducing itinerant holes induces ferromagnetic correlations, which persist for densities up to $x \approx 0.1$ (the region with positive Curie-Weiss temperature labeled by the thick blue line). However, magnetic ordering is observed for a smaller range of densities; the ordering temperature plunges when $x$ crosses $x_c$. Below $x_c$, the system is insulating. Top panel of Fig. 4 shows the itinerant hole density dependence of the localization length ($\xi$) and $\sqrt{A}$ that can be extracted from experiment (Extended Data

Fig. 4b and 8). Both $\xi$ and $\sqrt{A}$ exhibit a strong upward change upon approaching $x_c$. Above $x_c$, a heavy Fermi liquid emerges below the Kondo coherence temperature, which increases with increasing $x$ as one moves deeper inside the metallic phase. Figure 4 also shows that for a range of doping density (0.04 – 0.1), a heavy Fermi liquid with strong ferromagnetic fluctuations is realized.

The experimental phase diagram prompts many questions, including the nature of the metal-insulator transition, the microscopic mechanism for the ferromagnetic correlations at small $x$, and the relation between the different density-tuned phase transitions. The metal-insulator transition is likely driven by disorder. The resistance on the insulating side is well described by the Efros-Shklovskii variable-range hopping model [35,36], $R_{xx} \propto \exp(T^{-1/2})$, from which we evaluate the charge carrier localization length $\xi$ (Extended Data Fig. 8 and Methods). The localization length drops sharply as $x$ decreases, but remains about 20 times the moiré period even for the lowest densities studied (Fig. 4). The critical density ($2 \times 10^{11}$ cm$^{-2}$) is also consistent with the typical disorder densities in MoTe$_2$/WSe$_2$ moiré bilayers [23–25].

Multiple mechanisms can possibly induce ferromagnetic correlations between the local moments [8,9,37]. In MoTe$_2$/WSe$_2$ moiré bilayers, the Heisenberg-type antiferromagnetic exchange between nearest-neighbor moments, $J_H = 2t_{Mo}^2/U$ with $t_{Mo}$ denoting the hopping integral in the Mo layer, is very weak as suggested by the near-zero Curie-Weiss temperature at $x = 0$. The Kondo-type (antiferromagnetic) exchange between the local moments and itinerant holes, $J_K = 2t_\perp^2/(\frac{1}{\Delta} + \frac{1}{U-\Delta})$, is gate-tunable through the charge-transfer gap $\Delta$ ($t_\perp$ is the interlayer hopping integral) [5]. It has been shown that for $J_K \gg J_H$, Kondo coupling of local moments to dilute itinerant carriers, together with correlated hopping processes, can overcome the Heisenberg antiferromagnetic exchange to induce ferromagnetic order [8]. Another mechanism is the Ruderman-Kittel-Kasuya-Yosida (RKKY) interaction [38–40]. Ferromagnetic correlation between local moments is favored when the itinerant carrier density is low such that the corresponding Fermi wavelength $\sim(xn_M)^{-1/2}$ is larger than the local moment distance. The evolution from a ferromagnetic to antiferromagnetic correlation with further increase of density in Fig. 4 is consistent with this mechanism. Future studies are required to identify the dominant mechanism at different doping densities.

Our experiment shows that the metal-insulator transition, Kondo breakdown transition and onset of ferromagnetic ordering all occur remarkably around $x_c$. One possible scenario is that the Kondo destruction transition is induced by the disorder-driven metal-insulator transition. Below $x_c$, the Anderson-localized holes are less effective at Kondo-screening the local moments, and the ferromagnetic correlations between the local moments win over to induce ordering. Microscopically, even below $x_c$ Kondo singlets are expected to develop locally, driven in part by the lattice sites with stronger $J_K$ in the presence of disorders, and they compete directly with the local ferromagnetic order. Only as $x$ is increased beyond the critical value, the Kondo singlets develop globally (which can further suppress charge localization due to the large Fermi surface) and the ferromagnetic order completely disappears. We observe these competing tendencies in real-space in a parton mean-field study of the moiré Kondo model within a minimal model of quenched positional disorder (Extended Data Fig. 9, 10 and Methods). Using appropriate choice of model parameters, we can reproduce all the key features of the phase diagram in Fig. 4. However, future studies on higher quality samples are warranted to reveal the relation between the different density-tuned quantum phase transitions.

## Methods
### Device fabrication and magneto transport measurements
We fabricate dual-gated devices of MoTe$_2$/WSe$_2$ moiré bilayers by the layer-by-layer dry transfer method [41]. Details have been described in Ref. [5]. In short, we align the WSe$_2$ and MoTe$_2$ monolayer crystals according to their crystallographic orientations determined by the optical second-harmonic generation measurement [21,22]. The resulting bilayers are either 0- or 60-degree-aligned. We sort them according to their distinct electric-field dependences of resistance at $\nu = 1$ (Ref. [23,24]). In this study, two devices of 60-degree-aligned (AB-stacked) bilayers are studied. The results presented in the main text are from device 1. They are reproduced in device 2 (Extended Data Fig. 5). In both devices, both the top and bottom gates are made of hexagonal boron nitride (hBN) and few-layer graphite. The thickness of hBN is about 10 nm in the bottom gate and 4 nm in the top gate. Thin Pt (about 5 nm) is used to make contacts to the moiré bilayers.

We perform electrical measurements down to 300 mK in a closed-cycle $^4$He cryostat with a $^3$He insert (Oxford TeslatronPT) and down to 10 mK (lattice temperature) in a dilution refrigerator (Bluefors LD250). Low bias (0.2 - 1 mV) is used to avoid sample heating. A voltage pre-amplifier (100-MΩ impedance) is used to measure the sample resistance up to 10 MΩ. The standard low-frequency (10 - 20 Hz) lock-in technique is used. Finite longitudinal-transverse coupling occurs in our devices. We symmetrize and anti-symmetrize the measured $R_{xx}$ and $R_{xy}$ under positive and negative magnetic fields, respectively, to obtain the longitudinal and Hall resistances.

### Magneto optical measurements
We perform the reflective magnetic circular dichroism (MCD) spectroscopy down to 1.6 K in a closed-cycle $^4$He cryostat (Attocube, Attodry 2100). Details have been described in Ref. [42]. In short, a superluminescent diode centered approximately at 1070 nm is used as the light source and a liquid nitrogen-cooled InGaAs array sensor is used as the detector. A microscope objective (0.8 numerical aperture) focuses the light to a spot of about 1.5 μm in diameter on the devices. The incident light intensity is kept below 20 nW/μm$^2$ to avoid heating or photo-doping. To obtain the reflectance contrast spectrum, we employ a reference spectrum, which is measured on a heavily doped device and is featureless in the relevant spectral region. The MCD spectrum is defined as $\frac{I^+ - I^-}{I^+ + I^-}$, where $I^+$ and $I^-$ denote the reflection spectrum of the left and right circularly polarized light, respectively. The MCD is strongly enhanced near the attractive polaron resonance of MoTe$_2$. We integrate its absolute value within the spectral window of 1.114 - 1.132 eV to analyze the spin polarization at doping $\nu = 1 + x$.

**Analysis of the transport data**

The Kondo temperature $T^*$ is evaluated from the temperature-dependent resistance for $x \gtrsim 0.05$ (Fig. 1e). The Kondo temperature marks the temperature scale below which coherent charge transport develops in the Kondo lattice, and ideally, the sample resistance drops significantly. In our experiment, we observe a broad crossover feature (a peak or a hump). We extract $T^*$ following the procedure described in Ref. [5] from the first minimum of $|\frac{dR_{xx}}{dT}|$. The value corresponds to either the resistance peak temperature or the temperature below which the resistance decreases significantly.

**Determination of the charge carrier localization length**

The localization length $\xi$ is evaluated from the temperature-dependent resistance for $x < x_c$ (Ref. [43]). Below about 50 K, $R_{xx}$ scales with temperature according to the Efros-Shklovskii variable-range hopping model, $R_{xx} \sim \exp(T_0/T)^{1/2}$ (Extended Data Fig. 8). We extract $\xi$ from the characteristic temperature $T_0$ using $\xi = \frac{2.8e^2}{4\pi\varepsilon\varepsilon_0 k_B T_0}$, where $e$, $\varepsilon$, $\varepsilon_0$, and $k_B$ denote the elementary charge, the dielectric constant of the substrate ($\varepsilon \approx 5$), the vacuum permittivity, and the Boltzmann constant, respectively.

**Mean-field phase diagram of the Moiré Kondo lattice model with disorders**

We model the MoTe$_2$/WSe$_2$ moiré Kondo lattice using a simplified Hamiltonian

$$H = -\sum_{r,r' \in W} t_W(r-r') c^\dagger_{r,\alpha} c_{r',\alpha} + J_K \sum_{\substack{\langle r,r' \rangle \\ r \in Mo, r' \in W}} S_r \cdot s_{r'} + \sum_{r,r' \in Mo} J_{r,r'} S_r \cdot S_{r'} \quad (1)$$

where $S_r$ denotes the local moment associated with the Mo layer and $s_r = c^\dagger_{r,\alpha} \sigma_{\alpha,\beta} c_{r,\beta}/2$ is the spin density associated with the itinerant hole from the W layer. The hopping integral of the itinerant holes in the W layer, $t_W$, is assumed to fall off exponentially with distance. The antiferromagnetic Kondo exchange, $J_K (> 0)$, is present between all sets of nearest-neighbor Mo and W-sites. We also include in the Hamiltonian a small nearest-neighbor antiferromagnetic exchange between the local moments $J_{r,r'}$. In addition, we include the self-consistently generated longer-ranged RKKY interactions between the local moments due to the coupling to the itinerant holes. To account for the Anderson localization at low itinerant hole densities, we assume a random distortion of the moiré lattice positions by no more than 0.6% from the average value. Such distortions lead to randomness in the (intralayer) hole hopping and (interlayer) Kondo coupling, which are assumed to be uncorrelated.

Next, we introduce the fermionic parton representation for the local moments, $S_r = f^\dagger_{r,\alpha} \sigma_{\alpha,\beta} f_{r,\beta}/2$, with a constraint for the average occupation $\langle f^\dagger_{r,\alpha} f_{r,\alpha} \rangle / N = 1$ for a system with $N = L \times L$ sites. To study the competition between the ferromagnetic phase and the heavy Fermi liquid phase, we perform self-consistent mean-field (MF) calculations with variational parameters

$$\phi_{r,r'} = \frac{1}{2} J_K(r) \sum_{r' \in \text{nn}r, \sigma} \left\langle c^\dagger_{r',\sigma} f_{r,\sigma} \right\rangle, \quad (2)$$

$$M_r^{c,f} = \sum_\sigma \langle \sigma n_{r,\sigma}^{c,f} \rangle, \quad (3)$$

for the Kondo singlet and magnetism, respectively. In addition, localization properties are quantified by the inverse participation ratio (IPR), $\text{IPR}_\ell = \sum_{i=1}^N |\psi_\ell(i)|^4$, where $\psi_\ell(i)$ is the wavefunction associated with the $\ell$-th eigenstate at site $i$. The MF calculations are done for each disordered configuration. Note that the 120° Néel order is suppressed in our numerical calculations due to the small antiferromagnetic exchange.

The MF phase diagram obtained by averaging over 50 uncorrelated disordered configurations is shown in Extended Data Fig. 9. We indeed find a density-tuned Kondo breakdown transition concomitant with a metal-to-insulator transition at a critical density $x_c \approx 0.05$ for a specific choice of parameters. At low values of $x$, there is no macroscopic Kondo screening, while the self-consistent (ferromagnetic) RKKY interaction complemented with the Kondo coupling gives rise to long-range ferromagnetic ordering for the local moments. The W-holes are strongly localized as shown in the IPR result in Extended Data Fig. 9. As hole density is increased, the ferromagnetism is suppressed and the Kondo singlets develop globally. In this heavy Fermi liquid phase with a large Fermi surface, the W-holes are delocalized (Extended Data Fig. 9) because the localization is suppressed by a factor of $\sim 1/O(k_F \ell)$, where $k_F$ is the Fermi wavevector and $\ell$ is the mean-free path. Local competitions between the magnetism and the Kondo singlet can also be captured in the real-space snapshots across the phase transition (see Extended Data Fig. 10).

**Acknowledgement**
This work is supported by the Air Force Office of Scientific Research under award number FA9550-19-1-0390 (transport measurements), the National Science Foundation (Platform for the Accelerated Realization, Analysis, and Discovery of Interface Materials) under cooperative agreement nos. DMR-2039380 (sample and device fabrication) and DMR-1807810 (optical measurements). SK and DC are supported in part by a NSF CAREER grant (DMR-2237522) and a Sloan research fellowship to DC. The growth of the hBN crystals was supported by the Elemental Strategy Initiative of MEXT, Japan, and CREST (JPMJCR15F3), JST. We use the Cornell Center for Materials Research Shared Facilities supported through the NSF MRSEC programme (DMR-1719875) and of the Cornell NanoScale Facility, an NNCI member supported by NSF grant NNCI-2025233. We also acknowledge support from a David and Lucille Packard Fellowship (K.F.M.), Kavli Postdoctoral Fellowship (W.Z.) and Swiss Science Foundation Postdoc Fellowship (P.K.).

**Author contributions**
B.S. and W.Z. fabricated the devices. W.Z. and B.S. performed the electrical transport measurements and analyzed the data with the help of Z.H. and Y.Z.. W.Z., Z.T. and P.K. performed the optical measurements. S.K. and D.C. performed the theoretical analysis. K.W. and T.T. grew the bulk hBN crystals. J.S. and K.F.M. designed the scientific objectives and oversaw the project. All authors discussed the results and commented on the manuscript.

# Figures

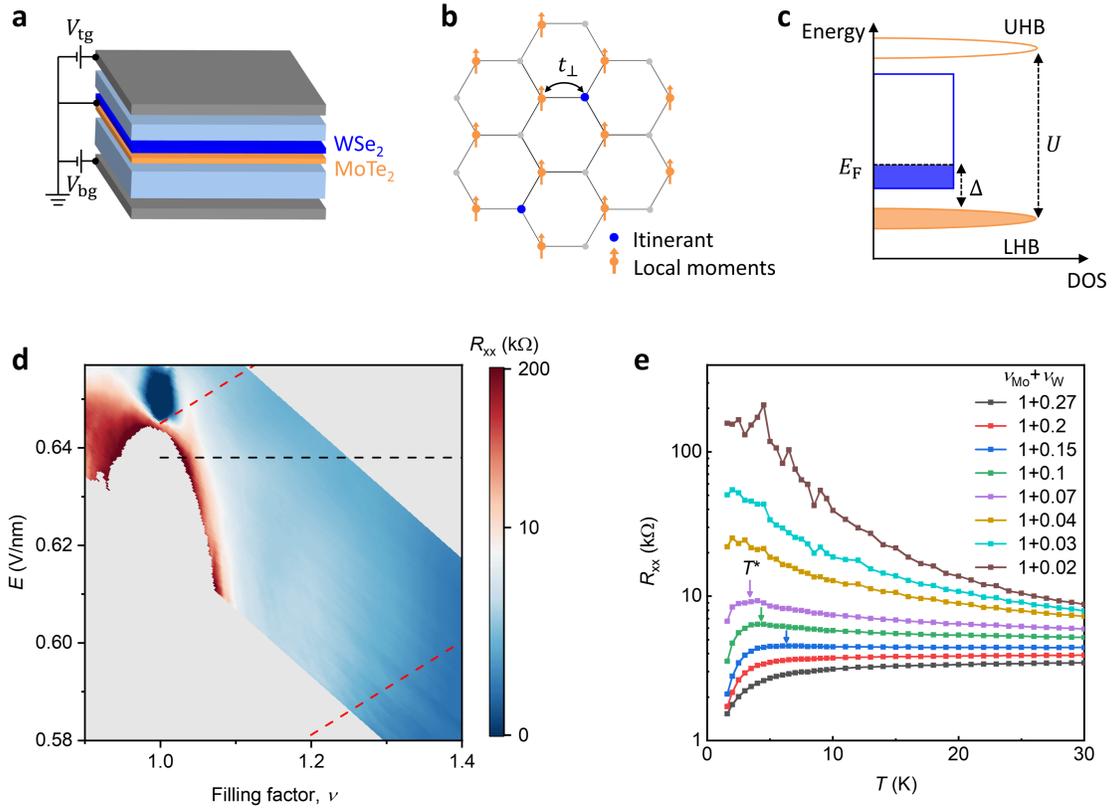

**Figure 1 | Gate-tunable moiré Kondo lattices**. **a,** Schematic of an AB-stacked MoTe$_2$/WSe$_2$ moiré bilayer controlled by two gate voltages ($V_{tg}$, $V_{bg}$). Both gates are made of hBN (light blue) and few-layer graphite electrodes (grey). **b,** The Wannier states of the two TMD layers form a staggered honeycomb moiré lattice with interlayer hopping integral $t_\perp$. The MoTe$_2$ layer is filled with one particle per moiré unit cell (orange) and forms a Mott insulator. The WSe$_2$ layer is filled with $x$ particles per unit cell (blue) on average and is weakly correlated. Ferromagnetic order is observed near the onset of a Kondo breakdown transition at small $x$. **c,** Schematic density of states (DOS) in the Kondo regime. The filled lower Hubbard band (LHB) and the empty upper Hubbard band (UHB) are separated by a Mott gap ($\sim U$). The weakly correlated WSe$_2$ band is between the LHB and UHB; $E_F$ is the Fermi energy; and $\Delta$ is the charge transfer gap. **d,** Longitudinal resistance $R_{xx}$ as a function of hole density in the bilayer, $\nu$, and out-of-plane electric field, $E$, at 10 mK. The red dashed lines mark the boundaries for a Kondo lattice region. The black dashed line corresponds to $E$ = 0.638 V/nm. **e,** Temperature dependence of $R_{xx}$ at varying hole densities ($x$) in the WSe$_2$ layer; $E$ is fixed along the black dashed line in **d**.

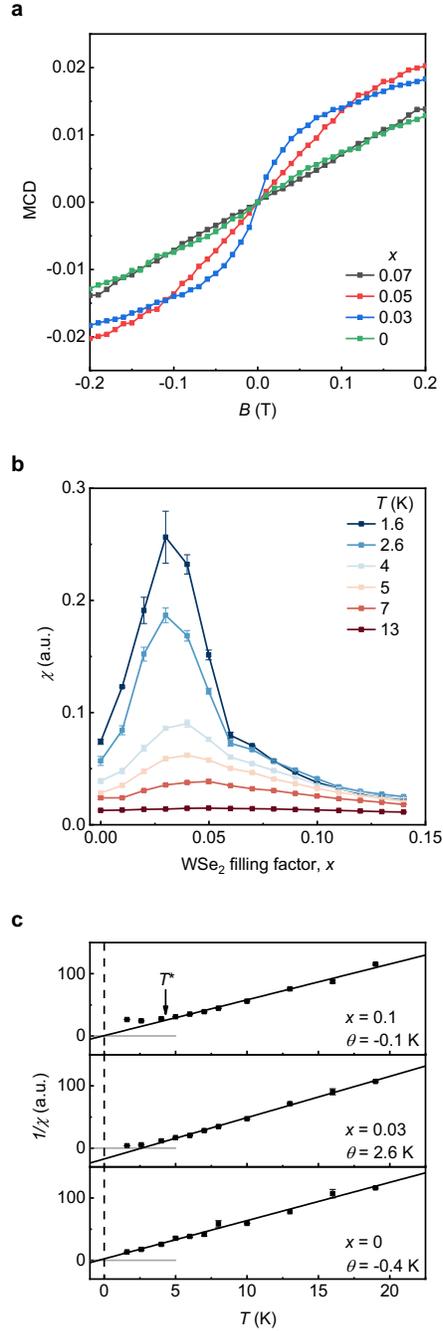

**Figure 2 | Magnetic correlations. a,** Magnetic-field dependence of MCD for representative itinerant hole densities at 1.6 K and $E = 0.638$ V/nm. **b**, Doping dependence of magnetic susceptibility $\chi$ at varying temperatures. **c**, Temperature dependence of inverse susceptibility (symbols) and Curie-Weiss fit (solid lines) with the best-fit Curie-Weiss temperature $\theta$. The three panels correspond to $x = 0.1$ (top), 0.03 (middle), and 0 (bottom). The vertical dashed line and horizontal grey lines denote, respectively, $T = 0$ and $1/\chi = 0$. The arrow in the top panel marks the Kondo temperature $T^*$. The error bars in **b, c** are obtained from the fitting uncertainty of the zero-field slope of MCD in **a**.

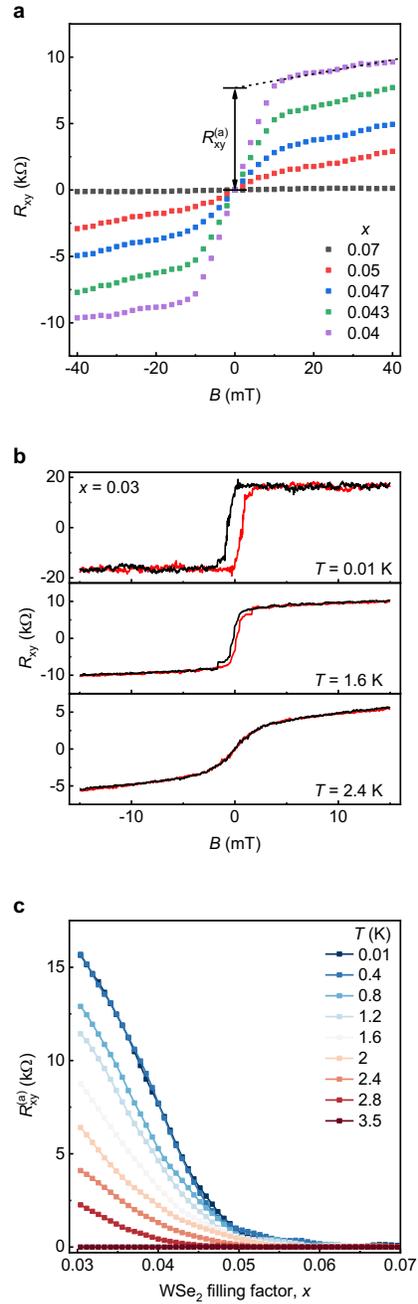

**Figure 3 | Anomalous Hall resistance and magnetic hysteresis. a**, Magnetic-field dependence of $R_{xy}$ at representative doping densities and 10 mK. The anomalous Hall resistance $R_{xy}^{(a)}$ is determined from the resistance jump at zero field from 0 to the projected ordinary Hall resistance by linear extrapolation from the data between 10 mT to 40 mT (black dotted line). **b**, Magnetic hysteresis at $x = 0.03$ at different temperatures. The hysteresis becomes indiscernible above about 2 K. **c**, Density dependence of $R_{xy}^{(a)}$ at varying temperatures. At 3.5 K, $R_{xy}^{(a)}$ vanishes for all densities.

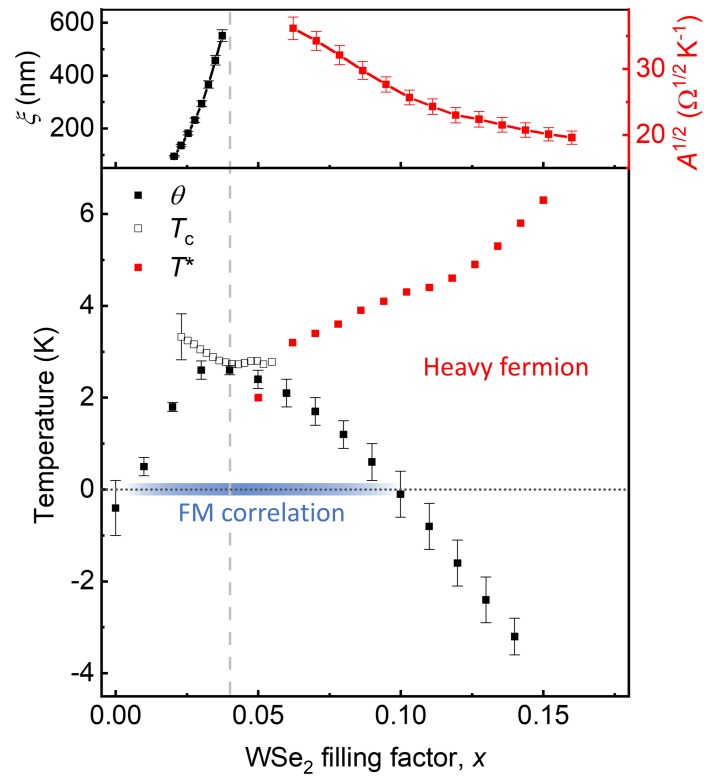

**Figure 4 | Phase diagram.** Density dependence of the localization length $\xi$, coefficient $A^{1/2}$, the Kondo temperature ($T^*$), magnetic ordering temperature ($T_c$) and Curie-Weiss temperature ($\theta$) from the transport and MCD measurements as described in the text. The thick blue line corresponds to the region with ferromagnetic correlations ($\theta > 0$) between the local moments. The dashed line denotes the critical density for the insulator-to-Fermi liquid transition.

# Extended Data Figures

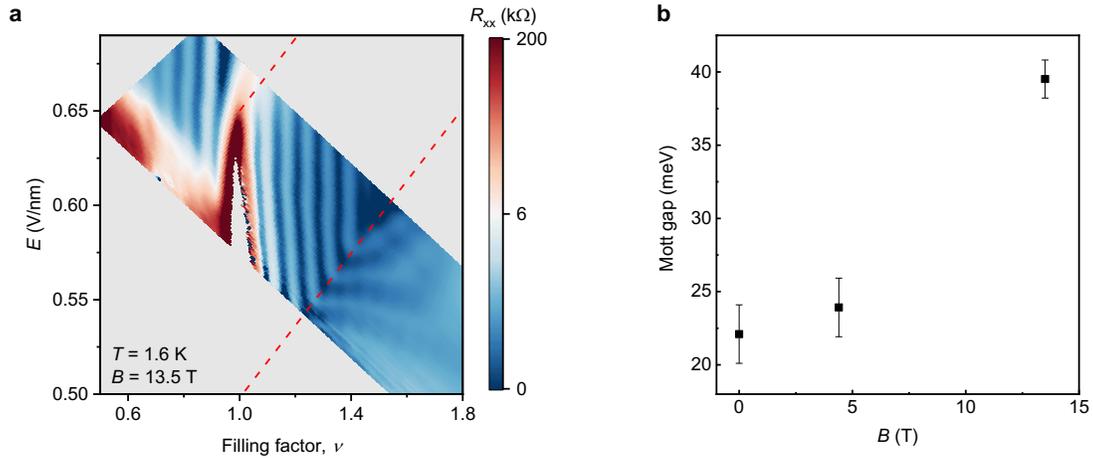

**Extended Data Figure 1 | Determination of the Kondo lattice region. a,** Longitudinal resistance $R_{xx}$ as a function of doping density in the bilayer, $\nu$, and out-of-plane electric field, $E$, at temperature $T = 1.6$ K and out-of-plane magnetic field $B = 13.5$ T. In the Kondo lattice region (bounded by the red dashed lines), the quantum oscillations from the WSe$_2$ layer shown as pronounced vertical stripes are independent of $E$. **b,** Magnetic field dependence of the Mott gap. The electric-field span of the Kondo lattice region multiplied by the interlayer dipole moment (0.26 $e \times$ nm) provides an estimate of the Mott gap $U = 40$ meV at 13.5 T. The zero field value (22 meV) is obtained from the optical reflection contrast measurement (see Ref. [5]).

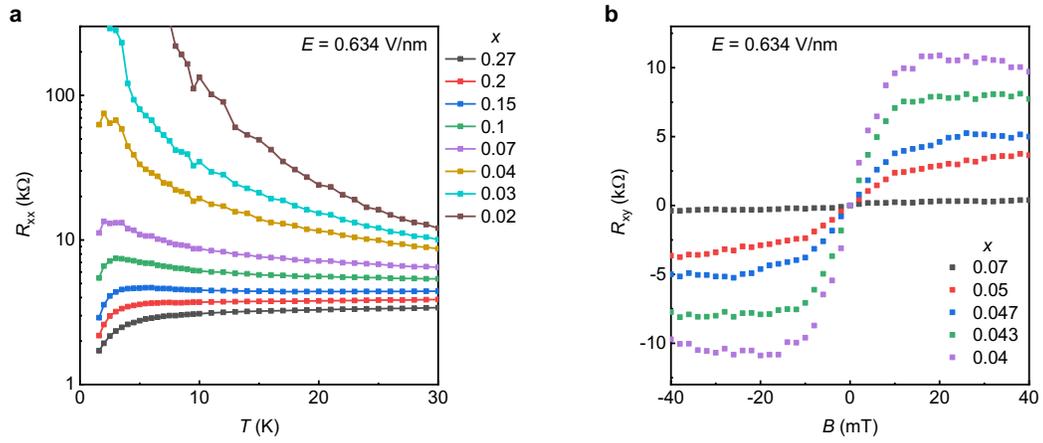

**Extended Data Figure 2 | Transport result at $E = 0.634$ V/nm. a**, Temperature dependence of $R_{xx}$ at varying doping densities in the Kondo lattice regime. **b**, Magnetic-field dependence of $R_{xy}$ at representative doping densities and 10 mK.

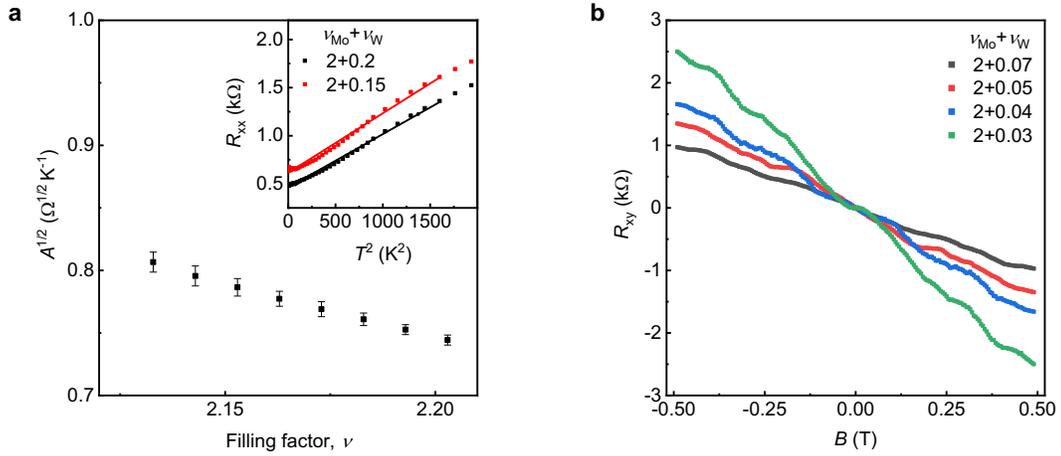

**Extended Data Figure 3 | Comparison to the case of $\nu = 2 + x$. a,** Doping dependence of the coefficient $A^{1/2}$ extracted from the temperature dependence of the longitudinal resistance $R_{xx}$. At low temperatures, $R_{xx}$ follows the dependence of $AT^2$ (inset, solid lines). The coefficient $A^{1/2}$ is one-to-two orders of magnitude smaller than for the case of $\nu = 1 + x$. **b,** Magnetic-field dependence of $R_{xy}$ at varying doping densities in the WSe$_2$ layer at 10 mK. The anomalous Hall response is absent for any $x$.

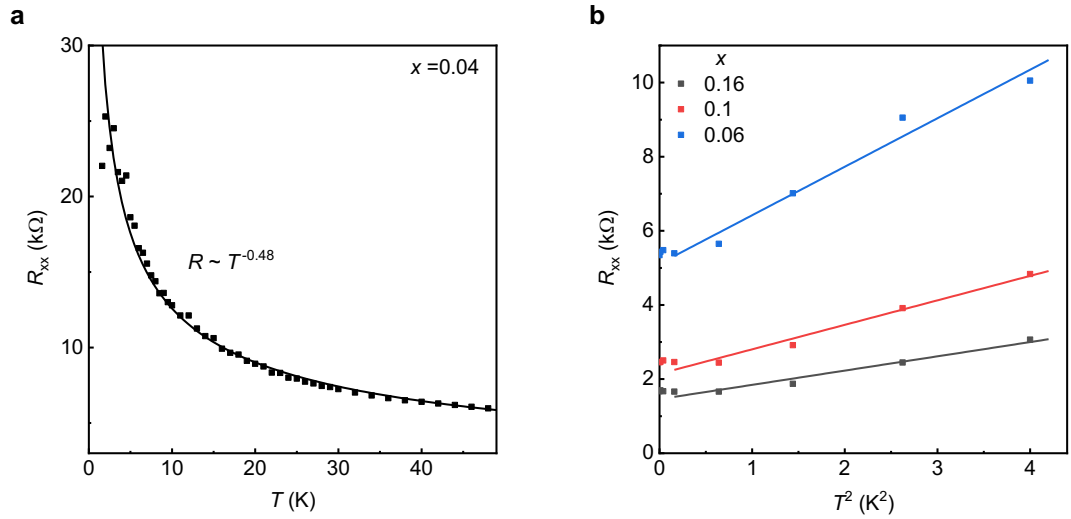

**Extended Data Figure 4 | Analysis of the temperature dependence of $R_{xx}$ at $E = 0.638$ V/nm ($\nu = 1 + x$). a**, At the critical density $x_c \approx 0.04$, $R_{xx}$ (symbols) follows a power-law temperature dependence, $R_{xx} \sim T^{-0.48}$ (solid line). **b**, Above the critical density, $R_{xx}$ scales linearly with $T^2$ at low temperatures. Solid lines are linear fits.

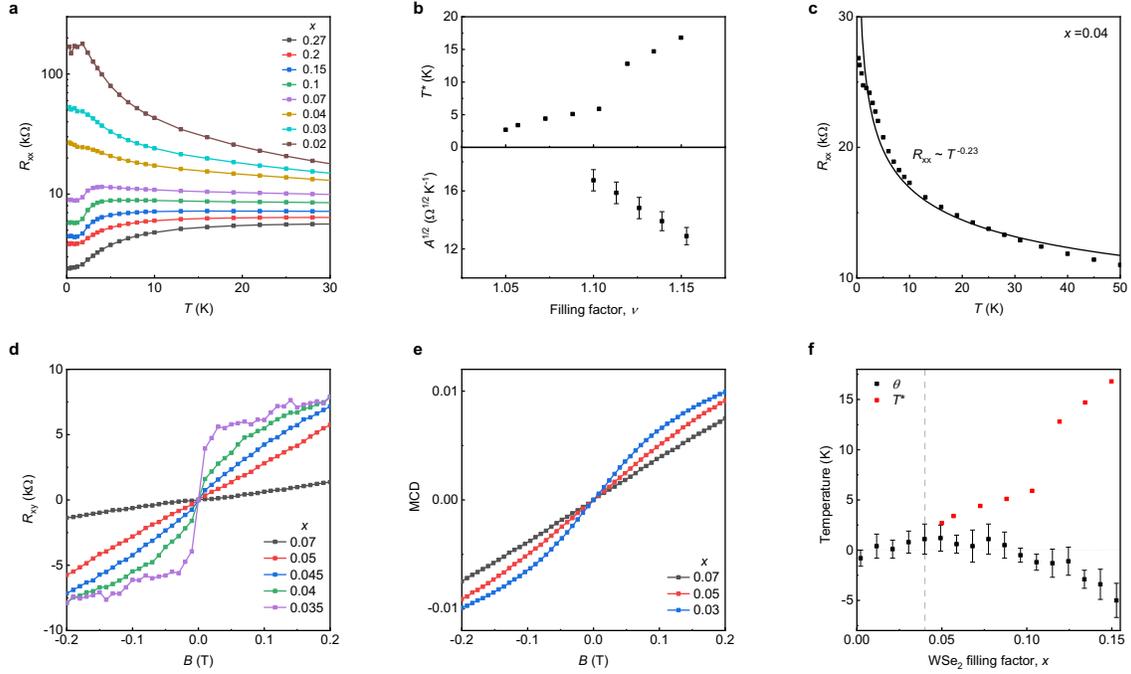

**Extended Data Figure 5 | Reproducible results in a second device. a**, Temperature dependent $R_{xx}$ at varying doping densities ($x$) in the Kondo lattice regime. **b**, Doping dependence of Kondo temperature $T^*$ and coefficient $A^{1/2}$ extracted from the data in **a** as described in the main text. **c**, At the critical density $x_c \approx 0.04$, $R_{xx}$ follows a power-law temperature dependence, $R_{xx} \sim T^{-0.23}$ (solid line). **d**, Magnetic-field dependence of $R_{xy}$ at representative doping densities in the WSe$_2$ layer at $T = 10$ mK. **e**, Magnetic-field dependence of MCD at representative doping densities in the WSe$_2$ layer at $T = 1.6$ K. **f**, Doping dependence of the Curie-Weiss and Kondo temperatures. Ferromagnetic correlations between local moments persist up to $x \approx 0.09$.

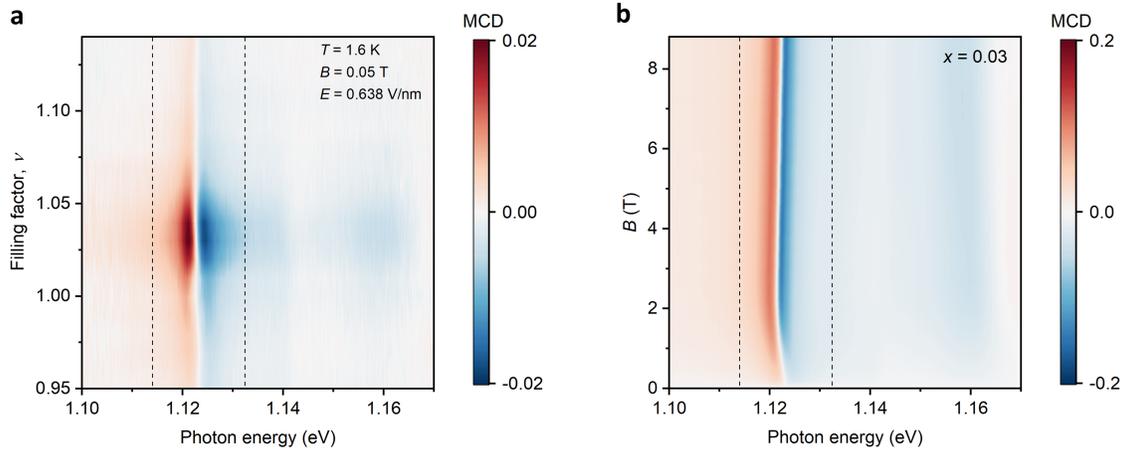

**Extended Data Figure 6 | MCD spectrum. a**, MCD spectrum centered around the exciton resonance of MoTe$_2$ as a function of doping density, $\nu$, in the bilayer at $T = 1.6$ K, $B = 0.05$ T and $E = 0.638$ V/nm. **b**, MCD spectrum as a function of magnetic field at $x = 0.03$ and $T = 1.6$ K. The absolute value of the MCD is integrated over the spectral window between the black dashed lines and presented as (integrated) MCD in the main text.

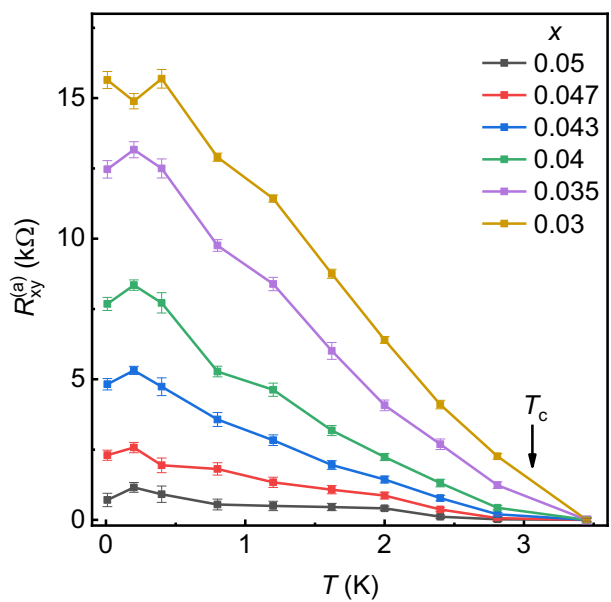

**Extended Data Figure 7 | Temperature dependence of anomalous Hall resistance.** The ferromagnetic order temperature $T_c$ is estimated as the temperature at which the anomalous Hall resistance $R_{xy}^{(a)}$ drops to 10% of its value at 10 mK. The arrow denotes $T_c$ for $x = 0.03$.

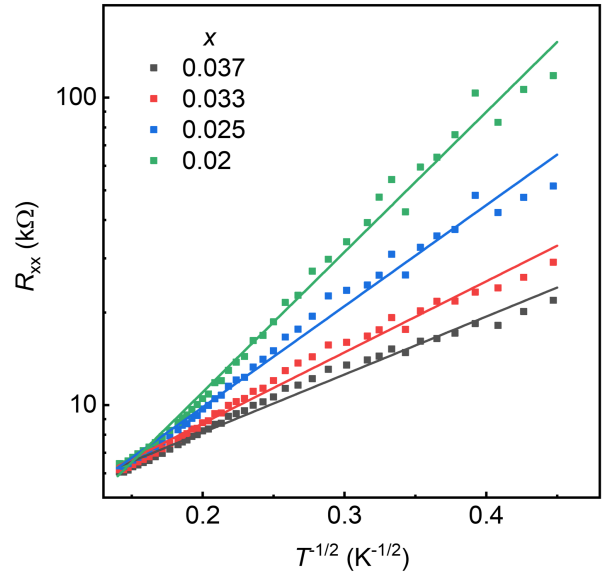

**Extended Data Figure 8 | Variable range hopping.** $R_{xx}$ (log scale) versus $T^{-1/2}$ at varying doping densities ($x$) in the Kondo lattice regime. Solid lines are linear fits, demonstrating Efros-Shklovskii variable range hopping.

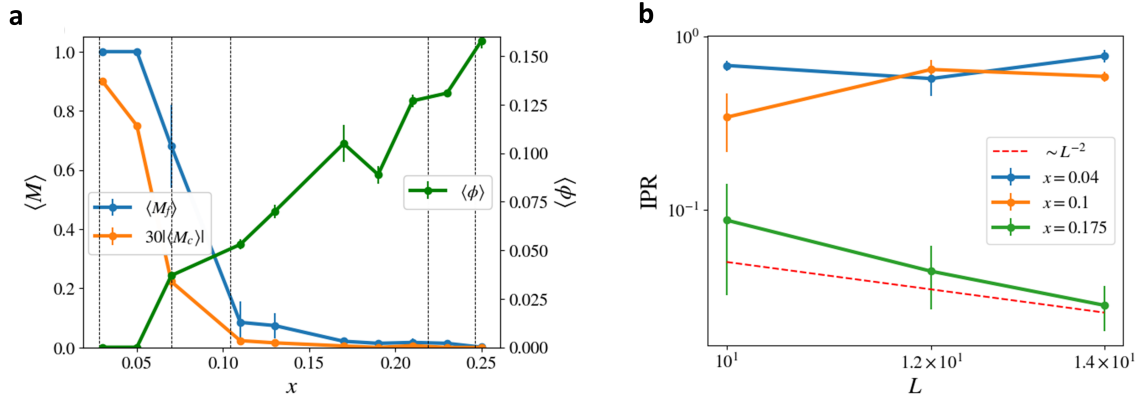

**Extended Data Figure 9 | Mean-field phase diagram and inverse participation ratio (IPR)**. **a,** MF phase diagram as a function of $x$ for a $L \times L$ (distorted) honeycomb lattice system ($L = 12$). The results are obtained by averaging over 50 uncorrelated disorder configurations. The Kondo-singlet MF parameter, $\langle|\phi|\rangle$, and the ferromagnetic component of the magnetization density for the Mo-hole (blue) and the W-hole (orange), $\langle M_f \rangle$, $|\langle M_c \rangle|$, respectively. The W-hole magnetism is multiplied by a constant numerical factor (30) to match the overall scale of the plot. We show in Extended Data Figure 10 the real-space snapshots for a fixed disorder configuration at the fillings marked by the vertical dashed lines. **b**, Disorder averaged IPR results as a function of system size. Three representative values of $x$ are chosen: $x = 0.04$ (FM phase), $x = 0.1$ (near the critical point), and $x = 0.175$ (heavy Fermi liquid phase).

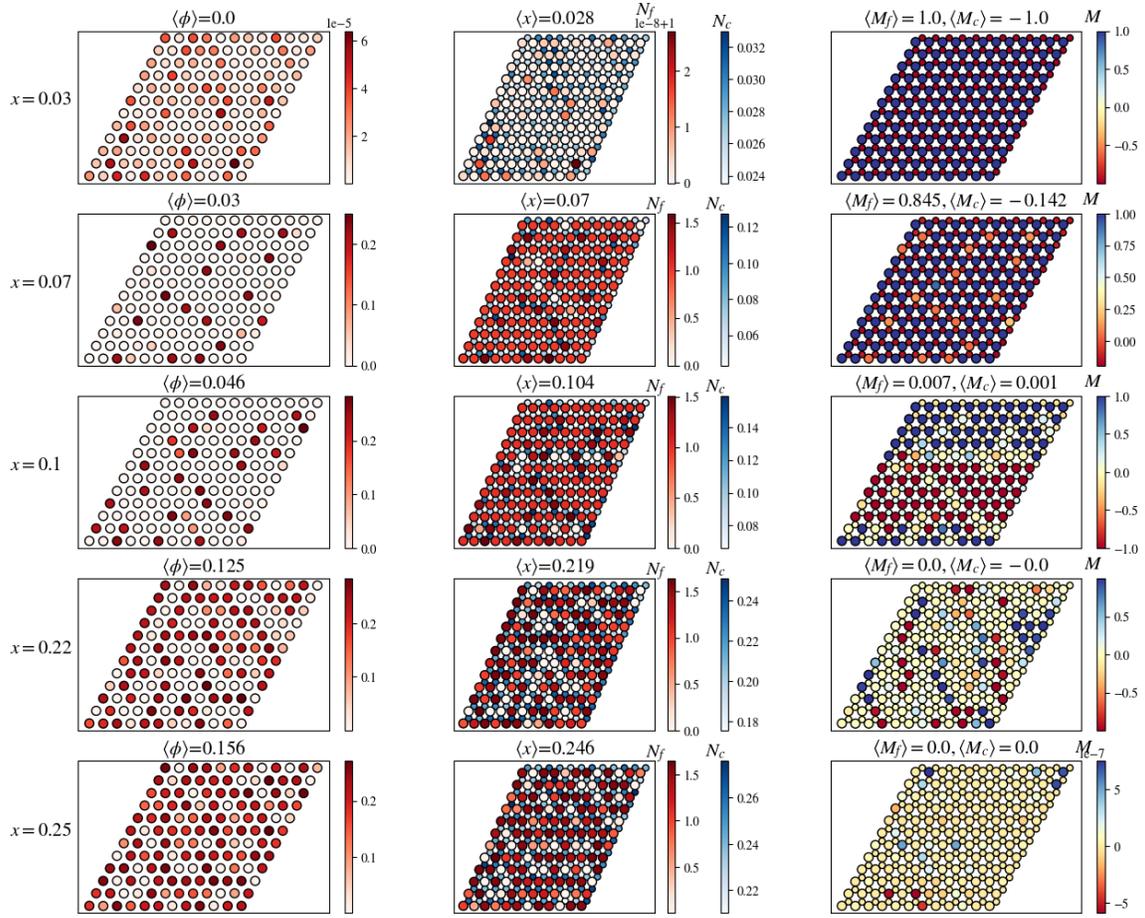

**Extended Data Figure 10 | Real-space snapshots across the phase transition.** Local expectation values of $|\phi|$ (left), $N_f, N_c$ (middle), and $M_f, M_c$ (right) for a disorder configuration at five representative values of $x$ (ascending order from top to bottom). At $x = 0.03$, both of the W holes and local moments show the long-range ferromagnetism. As $x$ is increased, local moments begin to give way to the Kondo singlet formation in an inhomogeneous fashion. At such sites, the local number constraint is violated. As $x$ exceeds the critical value, the Kondo singlets form globally and the ferromagnetism disappears.